\documentclass[twocolumn,prl,floatfix,citeautoscript,nofootinbib,superscriptaddress]{revtex4}
\usepackage{amsbsy}
\usepackage{latexsym,epsfig,graphicx}
\usepackage{dcolumn}
\usepackage{graphicx}
\usepackage{subfigure}
\usepackage{comment}
\usepackage{color}
\usepackage{bm}
\usepackage{mathrsfs}
\usepackage{amsfonts}
\usepackage{amsmath}
\usepackage{color}
\usepackage{amssymb}
\usepackage{xspace}
\usepackage{epstopdf}
\usepackage{tabularx}
\usepackage{longtable}
\usepackage[colorlinks=true, letterpaper=true, pdfstartview=FitV, linkcolor=red, citecolor=blue, urlcolor=blue]{hyperref}
\usepackage[normalem]{ulem}

\pdfoutput=1

\begin{document}

\title{Quantum squeezing and sensing with pseudo anti-parity-time symmetry}
\author{Xi-Wang Luo}
\affiliation{Department of Physics, The University of Texas at Dallas, Richardson, Texas
75080-3021, USA}
\author{Chuanwei Zhang}
\email{Chuanwei.Zhang@utdallas.edu}
\affiliation{Department of Physics, The University of Texas at Dallas, Richardson, Texas
75080-3021, USA}
\author{Shengwang Du}
\email{dusw@utdallas.edu}
\affiliation{Department of Physics, The University of Texas at Dallas, Richardson, Texas
75080-3021, USA}

\begin{abstract}
The emergence of parity-time ($\mathcal{PT}$) symmetry has greatly enriched
our study of symmetry-enabled non-Hermitian physics, but the realization of
quantum $\mathcal{PT}$-symmetry faces an intrinsic issue of unavoidable
symmetry-breaking Langevin noises. Here we construct a quantum pseudo-anti-$%
\mathcal{PT}$ (pseudo-$\mathcal{APT}$) symmetry in a two-mode bosonic system
without involving Langevin noises. We show that the spontaneous pseudo-$%
\mathcal{APT}$ symmetry breaking leads to an exceptional point, across which
there is a transition between different types of quantum squeezing dynamics,
\textit{i.e.}, the squeezing factor increases exponentially (oscillates
periodically) with time in the pseudo-$\mathcal{APT}$ symmetric (broken)
region. Such dramatic changes of squeezing factors and quantum dynamics near
the exceptional point are utilized for ultra-precision quantum sensing.
These exotic quantum phenomena and sensing applications can be
experimentally observed in two physical systems: spontaneous wave mixing
nonlinear optics and atomic Bose-Einstein condensates. Our work offers a
physical platform for investigating exciting $\mathcal{APT}$ symmetry
physics in the quantum realm, paving the way for exploring fundamental
quantum non-Hermitian effects and their quantum technological applications.
\end{abstract}

\maketitle

\emph{{\color{blue}Introduction}.---}Hermiticity and real eigenvalues of a
Hamiltonian are key postulates of quantum mechanics. While non-Hermitian
Hamiltonians emerged from the interaction with external environments
generally possess complex eigenspectra, they can exhibit entirely real
eigenvalues in the presence of parity-time ($\mathcal{PT}$) symmetry \cite%
{PhysRevLett.80.5243,Rep.70.947,nphys4323,s41563-019-0304-9,RevModPhys.88.035002,LFeng,science.aar7709}%
. When the non-Hermiticity parameter exceeds a critical value, known as
exceptional point (EP), the $\mathcal{PT}$-symmetry can be spontaneously
broken for the eigenstates, leading to a phase transition from the $\mathcal{%
PT}$-symmetric phase with purely real eigenvalues to the $\mathcal{PT}$%
-broken phase with complex conjugate eigenvalue pairs. In the past decade,
significant experimental and theoretical progress \cite%
{OL.32.002632,PhysRevLett.103.093902,nphys1515,nphys2927,PhysRevLett.110.083604,PhysRevLett.117.123601, PhysRevA.84.040101, PhysRevX.4.031042,ncomms6905, science.1258479,science.1258480, PhysRevLett.112.203901,PhysRevLett.117.110802}
has been made for exploring $\mathcal{PT}$-symmetry physics in various
physical systems (e.g., photonics, acoustics, ultracold atoms, etc.), which
generally utilize the control of linear gain/loss in classical wave systems.
However, an intrinsic issue for studying $\mathcal{PT}$-symmetry physics in
the quantum realm \cite%
{ncomms14154,ncomms4320,PhysRevLett.124.020501,PhysRevLett.123.180501,
GCGuo,Murch,JFDu} is that a $\mathcal{PT}$-symmetric Hamiltonian involving
linear gain/loss does not preserve the commutation relations of quantum
field operators, therefore Langevin noises, which break $\mathcal{PT}$
symmetry, must be included in quantum systems \cite{EPLScheel2018}. Two
experimental approaches to circumvent this issue for quantum $\mathcal{PT}$%
-symmetry include discarding quantum noise through post-selection
measurement \cite{Murch} and Hamiltonian dilation by embedding a
non-Hermitian Hamiltonian into a larger Hermitian system \cite{JFDu}.

Anti-$\mathcal{PT}$ ($\mathcal{APT}$) represents another non-Hermitian
symmetry with the Hamiltonian \textit{anticommuting} with $\mathcal{PT}$
operator (\textit{i.e.}, $\left\{ H_{APT},\mathcal{PT}\right\} =0$ instead
of commutation $\left[ H_{PT},\mathcal{PT}\right] =0$ for $\mathcal{PT}$%
-symmetry) and has recently attracted great interests \cite%
{PhysRevA.88.053810,nphys3842,PhysRevLett.113.123004,PhysRevA.96.053845,PhysRevLett.120.123902,PhysRevX.8.021066,science.aaw6259,PhysRevLett.123.193604,PhysRevA.99.063834,NJP.18}%
. Similar as $\mathcal{PT}$ symmetry, the spontaneous breaking of $\mathcal{%
APT}$ symmetry also leads to the emergence of EPs. Recent studies showed
that an $\mathcal{APT}$-symmetric system does not have to involve linear
gain/loss of classical fields, making it possible to realize a quantum $%
\mathcal{APT}$-symmetry without Langevin noises \cite%
{PhysRevLett.123.193604, NJP.18,PhysRevA.99.063834}. In this Letter, we
construct a quantum $\mathcal{APT}$-symmetry in a two-mode bosonic system,
where the dynamical Hamiltonian matrix is non-Hermitian and preserves the $%
\mathcal{APT}$-symmetry, while the second-quantized Hamiltonian is
Hermitian. In this sense, we name it a pseudo-$\mathcal{APT}$ symmetry. Our
main results are:

\textit{i}) The quantum pseudo-$\mathcal{APT}$-symmetry builds on coupling
Bose creation operator of one field with the annihilation operator of the
other field, yielding the Hermiticity of the second-quantized Hamiltonian
that does not involve Langevin noises. The spontaneous $\mathcal{APT}$
symmetry breaking across the EP for the dynamical Hamiltonian matrix yields
a transition from purely imaginary ($\mathcal{APT}$-symmetric) to purely
real ($\mathcal{APT}$-broken) eigenvalues, which is opposite to typical real
to imaginary transition for the $\mathcal{PT}$-symmetry.

\textit{ii}) Across the EP, the pseudo-$\mathcal{APT}$ symmetry and
quantized Hamiltonian yield a transition between different types of quantum
squeezing dynamics. Specifically, the two-mode squeezing factor oscillates
periodically with time in the pseudo-$\mathcal{APT}$-broken region,
increases linearly at EP, and grows exponentially in the pseudo-$\mathcal{APT%
}$-symmetric region. Optical field squeezed states have been widely studied
because of their fundamental interest (e.g., the implementation of EPR
paradox) as well as broad applications in quantum information processing
(e.g., continuous-variable quantum teleportation) and quantum metrology
(e.g., gravitational wave detection)~\cite{RevModPhys.77.513, Squeezed.light}%
. Here the connection between pseudo-$\mathcal{APT}$-symmetry transition and
different quantum squeezing dynamics is established.

\textit{iii}) The dramatic changes of quantum squeezing factors and dynamics
close to the EP make them ultra-sensitive to some parameters, thus can be
utilized to achieve ultra-precision quantum sensing. In contrast to quantum
sensing based on large squeezing factor~\cite%
{nphoton.2011.35,RevModPhys.90.035005}, here we focus on the $\mathcal{APT}$%
-broken region with weak squeezing that is usually undesirable in previous
experiments. We show that simple measurement schemes can reach the
sensitivity close to the quantum Cram\'{e}r-Rao bound given by the quantum
Fisher information \cite{nphoton.2011.35,RevModPhys.90.035005}, which
exhibits divergent feature as the EP is approached. The squeezing factor is
1 at the working points, therefore the ultra-precision sensitivity
originates from the pseudo-$\mathcal{APT}$ symmetry rather than squeezing or
entanglement.

\textit{iv}) We propose that the connection between the quantum pseudo-$%
\mathcal{APT}$ symmetry and the transition of squeezing dynamics as well as
the ultra-precision quantum sensors can be realized experimentally in
spontaneous wave mixing nonlinear optics and ultracold atomic Bose-Einstein
condensates (BECs). In the BEC case, we establish the connection between the
pseudo-$\mathcal{APT}$ transition and the well-known transition to dynamical
instability~\cite{BECbook}.

\emph{%
\color{blue}{Pseudo-$\mathcal{APT}$-symmetry and quantum
squeezing.---}}Consider a two-mode bosonic model described by the
second-quantized Hermitian Hamiltonian
\begin{equation}
\mathcal{H}=\delta \left( \hat{a}_{1}^{\dag }\hat{a}_{1}+\hat{a}_{2}^{\dag }%
\hat{a}_{2}\right) +i\kappa \left( \hat{a}_{1}^{\dag }\hat{a}_{2}^{\dag }-%
\hat{a}_{1}\hat{a}_{2}\right),  \label{SQH}
\end{equation}%
where $\hat{a}_{j}$ and $\hat{a}_{j}^{\dag }$ are the bosonic annihilation
and creation operators, and the detuning $\delta $ and coupling coefficient $%
\kappa $ are both real numbers. From Heisenberg equation, we obtain the
dynamical equation (we set $\hbar=1$) \cite{SM}%
\begin{equation}
i\partial _{t}\left(
\begin{array}{cc}
\hat{a}_{1}, & \hat{a}_{2}^{\dag }%
\end{array}%
\right) ^{T}=H_{APT}\left(
\begin{array}{cc}
\hat{a}_{1}, & \hat{a}_{2}^{\dag }%
\end{array}%
\right) ^{T},  \label{eq:bdg1}
\end{equation}%
with the non-Hermitian dynamical Hamiltonian matrix
\begin{equation}
H_{APT}=\delta \sigma _{z}+i\kappa \sigma _{x}=\left(
\begin{array}{cc}
\delta & i\kappa \\
i\kappa & -\delta%
\end{array}%
\right) .
\end{equation}%
$H_{APT}$ satisfies $\{H_{APT},\mathcal{PT}\}=0$, with the parity operator $%
\mathcal{P}=\sigma _{x}$ and the time-reversal complex conjugate operator $%
\mathcal{T}$. In the pseudo-$\mathcal{APT}$-symmetric region (\textit{i.e.},
the eigenstates of $H_{APT}$ are $\mathcal{PT}$-symmetric) $|\delta
|<|\kappa |$, $H_{APT}$ has two imaginary eigenvalues $\lambda _{\pm }=\pm
i\lambda _{0}$ with $\lambda _{0}=\sqrt{|\kappa ^{2}-\delta ^{2}|}$ \cite{SM}%
. While in the pseudo-$\mathcal{APT}$-broken region $|\delta |>|\kappa |$, $%
H_{APT}$ has two real eigenvalues $\lambda _{\pm }=\pm \lambda _{0}$. The
spontaneous symmetry breaking occurs at the EP $|\kappa |=|\delta |$, where $%
\lambda _{0}=0$. In the classical limit, the field operators $\hat{a}_{1}$
and $\hat{a}_{2}^{\dag }$ are replaced by complex numbers, and the model
reduces to the non-Hermitian system with $\mathcal{APT}$ symmetry \cite%
{NJP.18, PhysRevLett.123.193604}. In the quantum realm, we name it as pseudo-%
$\mathcal{APT}$ symmetry in the sense that $H_{APT}$ is non-Hermitian while $%
\mathcal{H}$ is Hermitian.

The field operators at time $t$ can be obtained from the dynamical equation (%
\ref{eq:bdg1}) as~\cite{SM}
\begin{equation}
\hat{a}_{j}(t)=A\hat{a}_{j}(0)+B\hat{a}_{\bar{j}}^{\dag }(0),
\label{fieldoperator}
\end{equation}%
where $\bar{j}$ represents the different mode number from $j$. In the pseudo-%
$\mathcal{APT}$-broken (symmetric) region, we have $A=\cos (\lambda _{0}t)-i%
\frac{\delta }{\lambda _{0}}\sin (\lambda _{0}t)$ ($A=\cosh (\lambda _{0}t)-i%
\frac{\delta }{\lambda _{0}}\sinh (\lambda _{0}t)$) and $B=\frac{\kappa }{%
\lambda _{0}}\sin (\lambda _{0}t)$ ($B=\frac{\kappa }{\lambda _{0}}\sinh
(\lambda _{0}t)$). $|A|^{2}-|B|^{2}=1$ in both regions and the bosonic
commutation relations $[\hat{a}_{j}(t),\hat{a}_{j^{\prime }}^{\dag }(t)]=[%
\hat{a}_{j}(0),\hat{a}_{j^{\prime }}^{\dag }(0)]=\delta _{jj^{\prime }}$ are
preserved without Langevin noises~\cite{SM}.

\begin{figure}[t]
\includegraphics[width=1.0\linewidth]{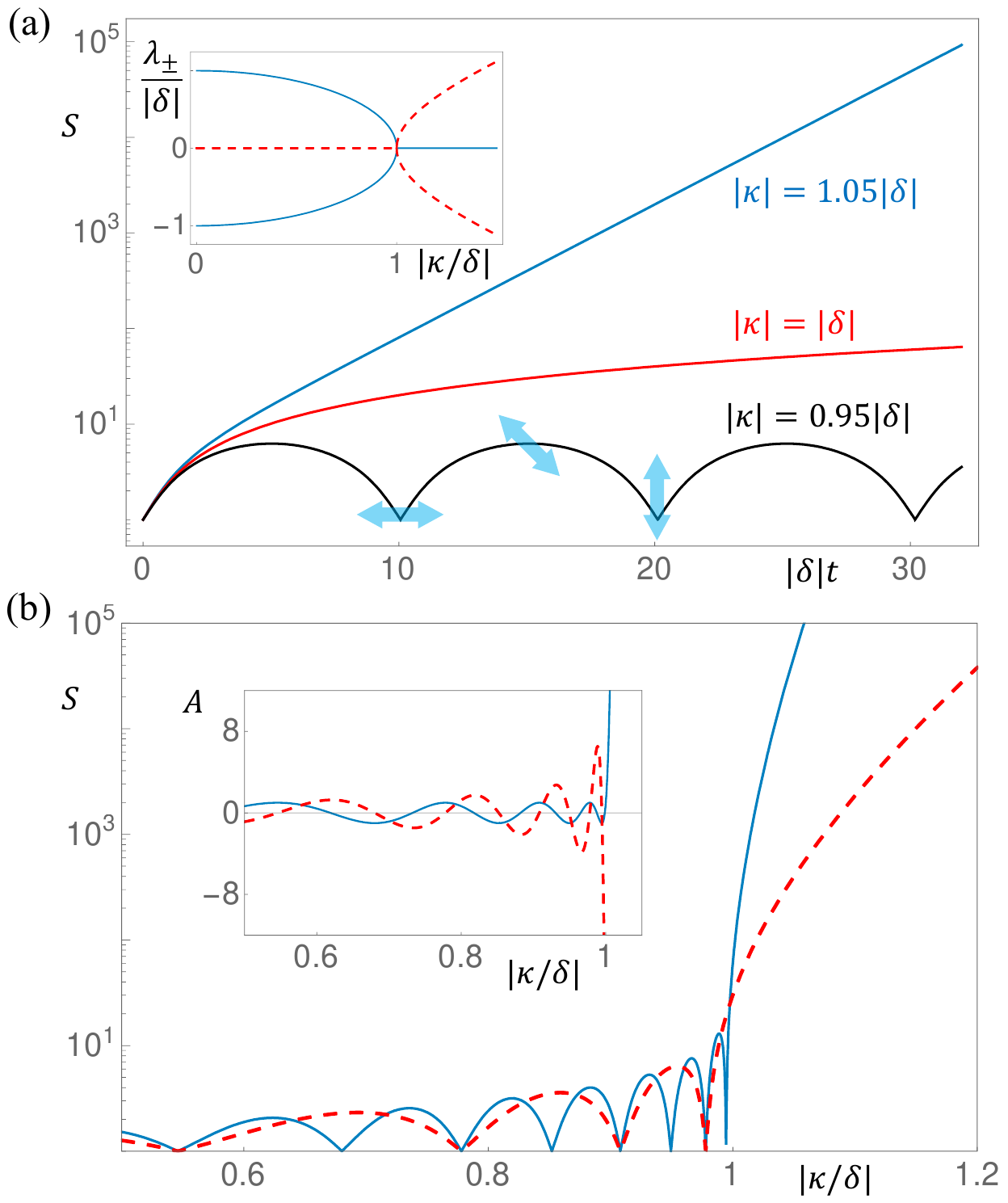}
\caption{Pseudo-$\mathcal{APT}$ symmetry induced quantum squeezing dynamics.
(a) The squeezing factor versus evolution time for different $\protect\kappa
$. $|\protect\kappa /\protect\delta |=0.95,1$ and $1.05$ correspond to
pseudo-$\mathcal{APT}$ broken, EP and pseudo-$\mathcal{APT}$ symmetric
regions, respectively. Blue arrows indicate the two-mode squeezing
directions for $\protect\kappa ,\protect\delta >0$. The inset shows the
eigenvalues of $H_{APT}$. Solid (dashed) lines are the real (imaginary)
parts. (b) Squeezing factor versus $\protect\kappa $ for different evolution
time. Solid (dashed) line corresponds to $|\protect\delta |t=30$ ($|\protect%
\delta |t=15$). Inset shows $A$ versus $|\protect\kappa /\protect\delta |$
with $|\protect\delta |t=30$. Solid (dashed) line is the real (imaginary)
part. $B=-{\protect\kappa }\mathrm{Im}[A]/{\protect\delta }$.}
\label{fig:1}
\end{figure}

The two-mode quantum squeezed states are generated from the terms $\hat{a}%
_{1}^{\dag }\hat{a}_{2}^{\dag }-\hat{a}_{1}\hat{a}_{2}$ in $\mathcal{H}$ and
can be characterized by the quadrature operators $\hat{X}_{j}(\varphi
,t)=[e^{-i\varphi }\hat{a}_{j}(t)+h.c.]/2$ of the two modes, which satisfy $%
\hat{X}_{1}(\varphi _{+},t)\pm \hat{X}_{2}(\varphi _{+},t)=S^{\pm 1}[\hat{X}%
_{1}(\varphi _{-},0)\pm \hat{X}_{2}(\varphi _{-},0)]$. Here $%
A=A_{0}e^{i\varphi _{A}}$ and $B=B_{0}e^{i\varphi _{B}}$ with the positive
amplitudes $A_{0}^{2}-B_{0}^{2}=1$, $S=A_{0}+B_{0}\geq 1$ is the two-mode
squeezing factor, $\varphi _{+}=(\varphi _{B}+\varphi _{A})/2$ is the
squeezing angle. The angle $\varphi _{-}=(\varphi _{B}-\varphi _{A})/2$ is
not important because the initial state is usually unentangled and isotropic
(e.g. the vacuum or coherent state).

Fig.~\ref{fig:1} shows the transition between different types of quantum
squeezing behaviors with the pseudo-$\mathcal{APT}$ symmetry starting from
an initial vacuum or coherent state. 
In the pseudo-$\mathcal{APT}$-symmetric region ($|\kappa /\delta |>1$), one
of the eigenmodes disappears after a long time evolution due to purely
imaginary $\lambda _{\pm }$, therefore $A_{0}\simeq B_{0}\simeq \frac{\kappa
}{2\lambda _{0}}e^{\lambda _{0}t}$ at the long time and $S\simeq \frac{%
\kappa }{\lambda _{0}}e^{\lambda _{0}t}$ grows exponentially. The squeezing
angle $\phi _{+}$ quickly changes from $\frac{1}{2}\mathrm{Arg}[\kappa ]$ to
its saturated value $\frac{1}{2}\mathrm{Arg}[\kappa \lambda _{0}-i\kappa
\delta ]$. In the pseudo-$\mathcal{APT}$-broken region, $S$ shows
oscillating behavior with a period $T=\pi /\lambda _{0}$, going back to 1
(non-squeezing) at $t=nT$ and reaching the maximum $S_{\max }=\sqrt{(|\delta
|+|\kappa |)/(|\delta |-|\kappa |)}$ at $t=(n+1/2)T$ ($n$ is an integer). In
each period starting from $S=1$, $\varphi _{+}$ changes from $\frac{1}{2}%
\mathrm{Arg}[\kappa ]$ to $\frac{1}{2}\mathrm{Arg}[-i\kappa \delta ]$ as $S$
increases to the maximum, and then to $\frac{1}{2}\mathrm{Arg}[-\kappa ]$ as
$S$ decreases to 1, as schematically illustrated in Fig.~\ref{fig:1}a. At
the EP, two eigenmodes coalesce to a single mode. We have $A=1-i\delta t$, $%
B=\kappa t$~\cite{SM}, and $S=\sqrt{1+\delta ^{2}t^{2}}+|\kappa |t$
increases linearly at long time $|\delta |t\gg 1$. $\varphi _{+}$ changes
monotonically from $\frac{1}{2}\mathrm{Arg}[\kappa ]$ to $\frac{1}{2}\mathrm{%
Arg}[-i\kappa \delta ]$. The results at the EP are consistent with the $%
|\kappa |\rightarrow |\delta |$ limit from both sides.

Since the squeezing behaviors change dramatically across the EP, the
dynamical quantum state at a given time should also be sensitive to the
system parameters $\delta $, $\kappa $ around the EP. In Fig.~\ref{fig:1}b,
we plot $S$ as a function of $|\kappa /\delta |$ at different times. We see $%
S$ oscillates with increasing amplitude and frequency as $|\kappa |$ is
approaching the EP. Near the EP, the squeezing factor (thereby the quantum
state) exhibits a sharp change around $S=1$ (\textit{i.e.}, for $\kappa $
satisfying $\lambda _{0}t=n\pi $), where the system returns to its initial
non-squeezing quantum state. The coefficients $A$ and $B$ show similar
oscillation behaviors as $S$ (the inset of Fig.~\ref{fig:1}b). Such critical
behavior around the EP can be utilized to implement ultra-precision quantum
sensing.

\emph{{\color{blue}Quantum sensing}.---}Precision measurements are long
pursued due to their vital importance in physics and many other sciences.
Quantum sensing, such as large squeezing factor state, quantum entanglement
\cite{nphoton.2011.35,RevModPhys.90.035005}, and phase-transition
criticality based sensors \cite%
{PhysRevA.88.021801,NJP.17,PhysRevA.93.022103,PhysRevLett.126.010502,PhysRevLett.124.120504,PhysRevX.8.021022}%
, utilize unique quantum phenomena for ultra-precision measurements. Recent
studies showed that the EPs of the $\mathcal{PT}$-symmetry can enhance the
optical sensing~in the classical region \cite{PRJ.396115}. Here we explore
ultra-precision quantum sensing enabled by the quantum pseudo-$\mathcal{APT}$%
-symmetry without Langevin noises.

We focus on the pseudo-$\mathcal{APT}$-broken region $|\kappa |<|\delta |$
(see~\cite{SM} for discussions on the region $|\kappa |>|\delta |$), which
is dynamical stable without the exponential growth of excitations. We
propose a simple scheme to measure $A$ and $B$ directly, which are sensitive
to the system parameters $\kappa $, $\delta $ and thus can be used to sense $%
\kappa $ and $\delta $. The sensing precision is at the same order as
quantum Cram\'{e}r-Rao bound set by the quantum Fisher information of the
quantum state, which shows divergent feature close to the EP.

\begin{figure}[t]
\includegraphics[width=1.0\linewidth]{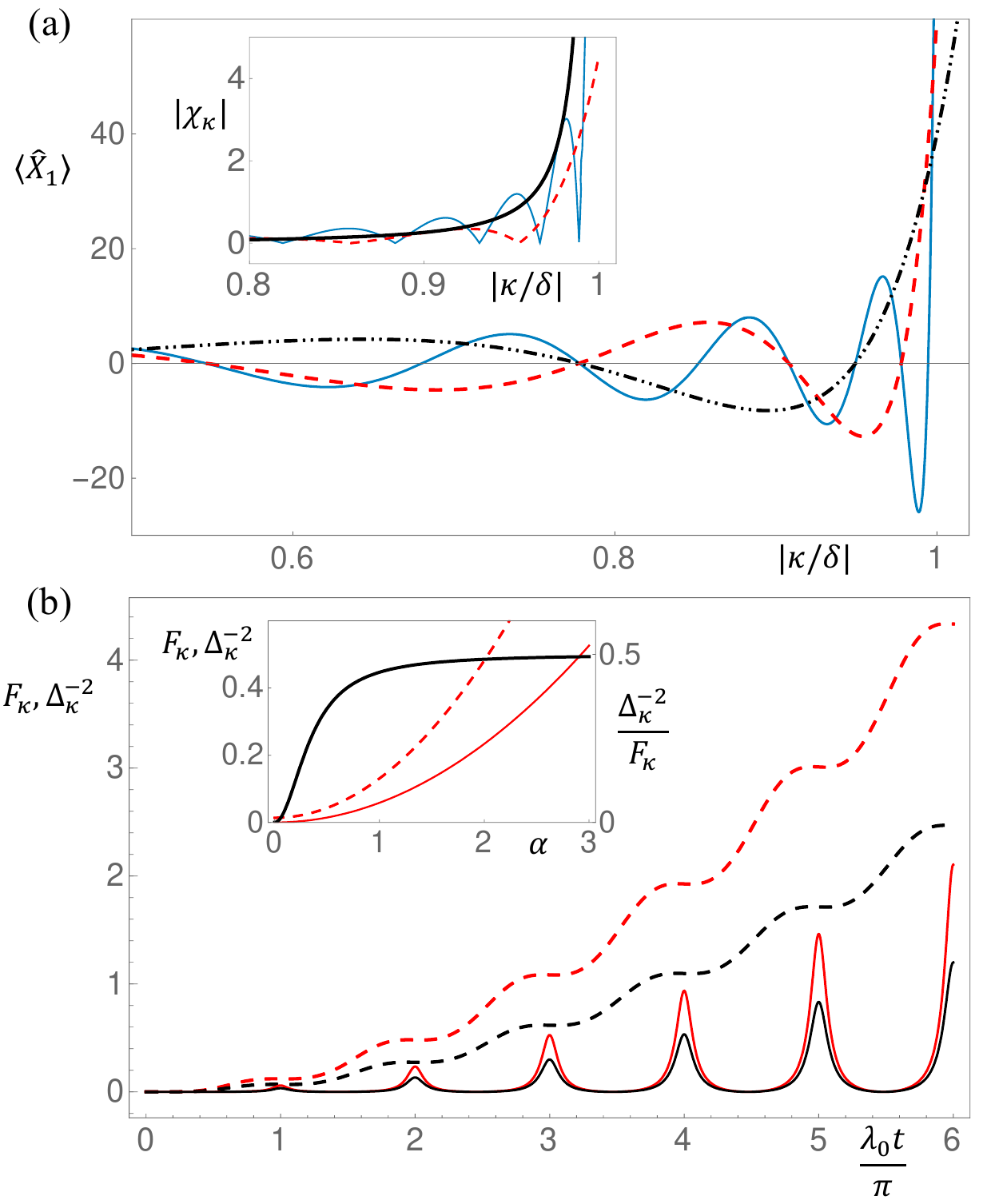}
\caption{Quantum sensing based on pseudo-$\mathcal{APT}$ symmetry. (a) The
quadrature $\langle \hat{X}_{1}\rangle $ versus $\protect\kappa $ at
different evolution time, with $\protect\alpha =2\cdot \mathrm{sign}(\protect%
\delta )$. $|\protect\delta |t=10,15$ and $30$ are shown by the dash-dotted,
dashed and solid lines, respectively. Inset shows the corresponding
susceptibility $|\protect\chi _{\protect\kappa }|$ (in unit of $10^{3}$),
with bold solid line showing the results with $\protect\kappa $-dependent
time satisfying $\protect\lambda _{0}t=2\protect\pi $. The working points
are located near $\langle \hat{X}_{1}\rangle =0$ (i.e., the maxima of $%
\protect\chi _{\protect\kappa }$). (b) The inverse variance $\Delta _{%
\protect\kappa }^{-2}$ (solid lines) of the observable and the quantum
Fisher information $F_{\protect\kappa }$ (dashed lines) as functions of
evolution times (in unit of $10^{7}$) with $\protect\alpha =2$. Red and
black lines are for $\protect\kappa /\protect\delta =0.95$ and $0.94$,
respectively. Local maxima of $\Delta _{\protect\kappa }^{-2}$ give the work
points $\protect\lambda _{0}t=n\protect\pi $. Inset shows the results as
functions of $\protect\alpha $ for $\protect\lambda _{0}t=2\protect\pi $ and
$\protect\kappa /\protect\delta =0.95$, with bold black line corresponding
to $\Delta _{\protect\kappa }^{-2}/F_{\protect\kappa }$.}
\label{fig:2}
\end{figure}

We consider a coherent initial state $|\psi _{0}\rangle =|\alpha _{1},\alpha
_{2}\rangle $ of two bosonic modes for the quantum sensor. After an
evolution time $t$, we perform the quadrature measurement $\hat{X}_{j}(0,t)$
of the final states using standard homodyne detection~\cite%
{Quantum.Measurement}, which give the mean value and variance
\begin{eqnarray}
\langle \hat{X}_{j}(0,t)\rangle _{\psi _{0}} &=&\mathrm{Re}[A\alpha
_{j}+B\alpha _{\bar{j}}^{\ast }] \\
\ [\Delta \hat{X}_{j}(0,t)]^{2} &=&\frac{1}{4}(A_{0}^{2}+B_{0}^{2}),
\end{eqnarray}%
with $\bar{j}\neq j$. Therefore we can determine $A$ and $B$ from the
measurement results for the estimation of $\kappa $ or $\delta $. Without
loss of generality, we choose $\kappa \delta >0$
and set $\alpha _{2}=-i\alpha _{1}\equiv \alpha $ (different choices of
parameters $\alpha _{i}$ give similar results, which do not affect the
sensing precision). In Fig.~\ref{fig:2}a, we plot the observable $\langle
\hat{X}_{1}(0,t)\rangle _{\psi _{0}}=\alpha \sin (\lambda _{0}t)\frac{\kappa
+\delta }{\lambda _{0}}$ as a function of $\kappa $ with fixed $\delta $ and
$t$, which possesses strong and fast oscillation close to the EP. Such
oscillation becomes more pronounced as the evolution time increases.

The measurement of the change of $\langle \hat{X}_{j}(0,t)\rangle _{\psi
_{0}}$ with $\kappa $ gives the susceptibility $\chi _{\kappa }(t)\equiv
\partial _{\kappa }\langle \hat{X}_{1}(0,t)\rangle _{\psi _{0}}$
which exhibits divergent feature close to the EP $\kappa \rightarrow \delta $
(\textit{i.e.}, $\lambda _{0}\rightarrow 0$). Similar results apply to the
susceptibility with $\delta $. In Fig.~\ref{fig:2}a, we plot $\chi _{\kappa
} $ for a fixed evolution time as well as the sensor working point $t=nT$.
The later one possesses a divergent scaling $\chi _{\kappa }(nT)=-\alpha
\frac{\kappa (\kappa +\delta )n\pi }{\lambda _{0}^{3}}\sim \lambda _{0}^{-3}$%
. Longer evolution time is required for smaller $\lambda _{0}$ to observe
such divergence. Note that the eigenvalues of $H_{APT}$ have a square-root
splitting with divergent sensitivity $\partial _{\kappa }\lambda _{0}=-\frac{%
\kappa }{\lambda _{0}}$ close to the EP. In addition, the eigenmodes are not
orthogonal and coalesce at the EP, which are responsible for the factor ${%
\lambda _{0}^{-1}}$ in the final Bosonic field $\hat{a}_{j}(t)$. Together
they give the divergent scaling $\chi _{\kappa }(t)\sim \lambda _{0}^{-3}$
for an evolution time $t\sim {\lambda _{0}}^{-1}$.

The precision of the parameter estimation of $\kappa $ is given by the
variance $\Delta _{\kappa }^{2}=[\Delta \hat{X}_{1}]^{2}/\chi _{\kappa }^{2}$%
, and the performance of the sensing scheme can be evaluated by comparing
the inverse variance $\Delta _{\kappa }^{-2}$ with the quantum Fisher
information $F_{\kappa }$, whose inverse gives the ultimate precision for
quantum sensing, \textit{i.e.}, reaching quantum Cram\'{e}r-Rao bound $%
\Delta _{\kappa }^{-2}\leq F_{\kappa }$ (optimal measurement is usually
required). For the coherent initial state $|\psi _{0}\rangle $,
\begin{eqnarray}
F_{\kappa }(t) &=&4A_{0}^{4}|\partial _{\kappa }\frac{B}{A^{\ast }}%
|^{2}+4(A_{0}^{2}+B_{0}^{2})\sum_{j}|\partial _{\kappa }\langle \hat{a}%
_{j}(t)\rangle _{\psi _{0}}|^{2}  \notag \\
&&-16\Re \lbrack A^{\ast }B^{\ast }\partial _{\kappa }\langle \hat{a}%
_{1}(t)\rangle _{\psi _{0}}\partial _{\kappa }\langle \hat{a}_{2}(t)\rangle
_{\psi _{0}}]  \label{eq:QFI}
\end{eqnarray}%
after the evolution time $t$ \cite{SM}. In Fig.~\ref{fig:2}b, we compare $%
\Delta _{\kappa }^{-2}$ with $F_{\kappa }$ at different evolution times. We
see that $\Delta _{\kappa }^{-2}$ has some narrow peaks when $\kappa $
satisfies $t=nT$ (\textit{i.e.}, the variance $\Delta _{\kappa }$ reaches
the minimum) for fixed $\delta $ and $|\kappa |\lesssim |\delta |$, while $%
F_{\kappa }(t)$ smoothly increases with $\kappa $ and takes larger values
for all $\kappa $ near the EP. At working points $t=nT$, $F_{\kappa }(nT)=%
\left[ 8+\frac{4\kappa ^{2}}{\alpha ^{2}(\kappa +\delta )^{2}}\right] \chi
_{\kappa }^{2}\sim \lambda _{0}^{-6}$, while $\Delta _{\kappa
}^{-2}(nT)=4\chi _{\kappa }^{2}\simeq 0.5F_{\kappa }$ for coherent
amplitudes $\alpha $ that are not too weak (e.g., $\alpha \sim 2$), as shown
in Fig.~\ref{fig:2}b. During the evolution, the number of bosonic
excitations $N\sim \lambda _{0}^{-2}$, therefore $\Delta _{\kappa }^{-2}\sim
N^{2}t^{2}$, which is at the same order as the Heisenberg limit.

Notice that at the working points $t=nT$, the squeezing factor $S=1$,
therefore the ultra-precision sensitivity originates from the pseudo-$%
\mathcal{APT}$ symmetry breaking, making our scheme distinct from
traditional quantum sensors based on large squeezing factors.

\begin{figure}[t]
\includegraphics[width=1.0\linewidth]{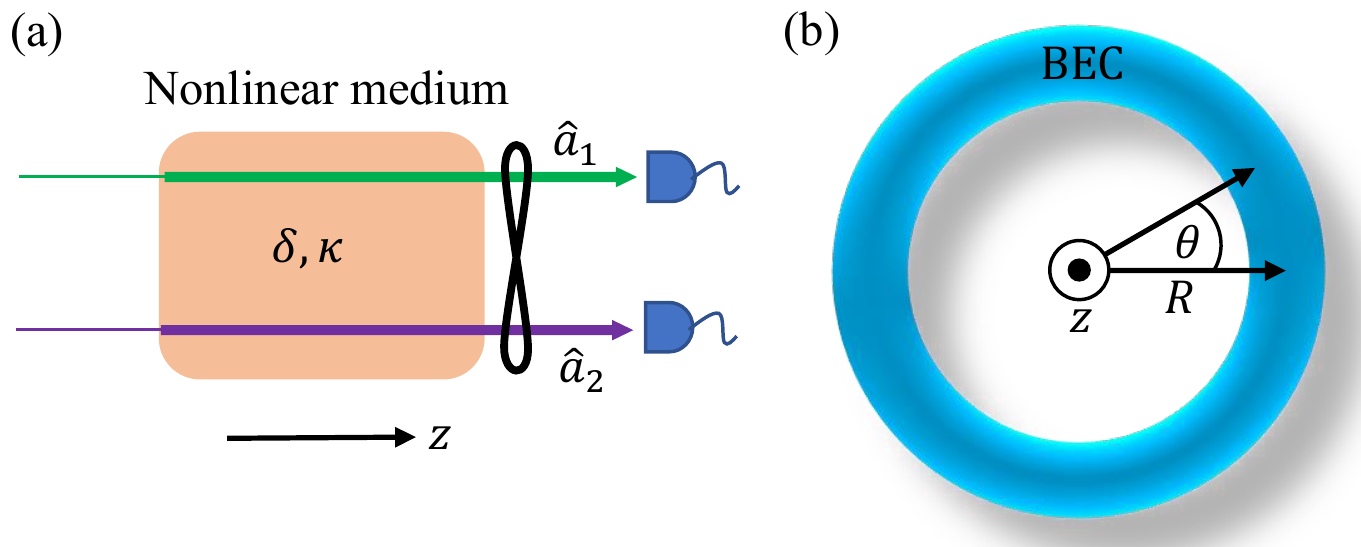}
\caption{Experimental implementations. (a) Schematic optical setup for
realizing pseudo-$\mathcal{APT}$ symmetry physics through nonlinear
spontaneous four-wave mixing. (b) Experimental realization based on a cold
atomic BEC in a ring trap.}
\label{fig:3}
\end{figure}

\emph{{\color{blue}Experimental realization.---}}The quantum pseudo-$%
\mathcal{APT}$ symmetry physics can be realized in quantum optics systems or
atomic BECs. In the quantum optics implementation, we can utilize nonlinear
wave mixing such as spontaneous parametric down conversion (SPDC) \cite%
{SPDC01, SPDC02} and spontaneous four-wave mixing (SFWM) \cite{SFWM01,
SFWM02} with carefully designed parameters. A schematic illustration of the
optical setup is shown in Fig.~\ref{fig:3}a and more detailed studies are
provided in~\cite{SM}. In SFWM, two quantum modes ($\hat{a}_{1,2}$)
copropagate along the $z$ direction in the nonlinear optical medium and are
coupled through a nonlinear coupling coefficient $\kappa $ which can be
tuned by changing two additional pump lasers' intensity and frequency, as
well as the material properties. The parameter $\delta $ is associated with
the phase mismatching $\delta =-\Delta k/2$ depending on laser frequency and
propagation direction, where $\Delta k=(\mathbf{k}_{1}+\mathbf{k}_{2}-\sum
\mathbf{k}_{\text{pump}})\cdot \hat{\mathbf{z}}$ with $\mathbf{k}_{j}$ being
the wave vectors and $\hat{\mathbf{z}}$ the unit vector. In this system, the
propagation along the $z$ direction simulates the time evolution in our
theoretical model (\textit{i.e.}, $t=z$). In the quantum sensing, the final
output quantum fields of two modes from the nonlinear medium will be
measured using standard homodyne detection, yielding the mean value and
variance of quadratures of two modes.

In the atomic implementation, we consider a BEC in a ring dipole trap (as
shown in Fig.~\ref{fig:3}b) with a strong confinement along $z$ and radial
directions, which can be realized by Laguerre-Gaussian lasers~as
demonstrated in recent experiments \cite{PhysRevA.74.023617,nature12958}.
The dynamics are reduced to one dimension (\textit{i.e.}, the azimuthal
angle $\theta $). 
We can expand the BEC field operator in the angular momentum space as~\cite%
{PhysRevB.53.9341,RevModPhys.77.187}
\begin{equation}
\hat{\Psi}(\theta ,t)=e^{-i\mu t-i\pi /4}\Phi (t)+e^{-i\mu t}\sum_{n\neq 0}%
\hat{\psi}_{n}(t)\frac{e^{in\theta }}{\sqrt{2\pi }},  \label{eq:field}
\end{equation}%
where $\Phi $ is the condensate wave function ($\Phi $ is real initially and
$\pi /4$ is a gauge choice), $\mu =-g|\Phi |^{2}-g\sum_{n}\langle \psi
_{n}^{\dag }\psi _{n}\rangle /2\pi $ is the chemical potential, and $g$ is
the interaction strength ($g>0$ corresponds to attractive interaction). The
quantum excitation operators $\hat{\psi}_{n}(t)$ satisfy the Bogoliubov
equation~\cite{SM}
\begin{equation}
i\partial _{t}\left(
\begin{array}{c}
\hat{\psi}_{n} \\
\hat{\psi}_{-n}^{\dag }%
\end{array}%
\right) =\left(
\begin{array}{cc}
\delta _{n} & i\kappa \\
i\kappa ^{\ast } & -\delta _{n}%
\end{array}%
\right) \left(
\begin{array}{c}
\hat{\psi}_{n} \\
\hat{\psi}_{-n}^{\dag }%
\end{array}%
\right) ,  \label{eq:bdg2}
\end{equation}%
which shares the same form as Eq.~\ref{eq:bdg1}. Here $\delta
_{n}=n^{2}E_{1}-{g}|\Phi |^{2}$, $\kappa ={g}\Phi ^{2}$, and $E_{1}=\frac{1}{%
2mR^{2}}$ is the kinetic energy of the first excited state along the ring
with radius $R$. At $g=0$, the quantum pseudo-$\mathcal{APT}$ symmetry is
broken for all $n$. As $g$ increases, the BEC becomes dynamical unstable
when $2g|\Phi |^{2}>E_{1}$ (i.e., $|\kappa |>|\delta _{1}|$), where the
pseudo-$\mathcal{APT}$ symmetry is restored for $n=1$, and the excitation
number and squeezing factor grow exponentially~\cite%
{PhysRevA.65.033605,PhysRevA.68.043625}. For the quantum
sensing, we can first prepare the BEC to its ground state with $g\simeq 0$.
The initial coherent state for $n=\pm 1$ can be generated by Raman process
with Laguerre-Gaussian beams carrying $\pm 1$ orbital angular momentum. Then
we increase $g$ to the working point near the EP (\textit{i.e.}, $2g|\Phi
|^{2}=E_{1}$). The quadratures $\hat{X}_{n}$, which is proportional to the
visibility of the density modulation along $\theta $, can be measured by
density imaging. The performance of the sensor is very similar as that shown
in Fig.~\ref{fig:2} \cite{SM}. We want to point out that the Kerr interaction of photons as well as the
interaction between excitations of the BEC are extremely weak,
which can hardly affect the finite duration dynamics of interest~\cite{SM}.

\emph{\color{blue}{Conclusion.---}}In summary, we construct a quantum pseudo-%
$\mathcal{APT}$ symmetry without Langevin noises and show that its
transition across EP yields a dramatic change of quantum squeezed dynamics.
The divergent sensitivity of squeezed states close to the EP can be utilized
for ultra-precision sensing approaching the quantum limit. The experimental
realization of such quantum pseudo-$\mathcal{APT}$ symmetry in nonlinear
quantum optical wave mixing and ultracold atomic BECs will provide realistic
platforms for studying quantum non-Hermitian physics and its quantum sensing
applications. The two-mode quantum pseudo-$\mathcal{APT}$ symmetry can also
be generalized to a multi-mode system supporting higher-order EPs~\cite%
{nature23280,s41467-019-08826-6}, which may lead to novel symmetry breaking
physics and even higher sensing precision.

\begin{acknowledgments}
\textbf{Acknowledgments.---}X.W.L. and C.Z. acknowledge support from NSF
(PHY-2110212), ARO (W911NF17-1-0128), and AFOSR
(FA9550-20-1-0220,FA9550-22-1-0043). S.D. acknowledges support from DOE
(DE-SC0022069).
\end{acknowledgments}

\newpage

\begin{widetext}
\setcounter{figure}{0} \renewcommand{\thefigure}{S\arabic{figure}} %
\setcounter{equation}{0} \renewcommand{\theequation}{S\arabic{equation}}

\section{Supplementary Materials}
\setcounter{figure}{0} \renewcommand{\thefigure}{S\arabic{figure}} %
\setcounter{equation}{0} \renewcommand{\theequation}{S\arabic{equation}}

\subsection{S1. Derivation of the dynamical equation}

As we discussed in the main text, the $\mathcal{APT}$-symmetric dynamical
equation is obtained from a second-quantized Hamiltonian which is Hermitian,
with pair generation of particles. The second-quantized Hamiltonian reads
\begin{equation}
\mathcal{H}=\delta \left[ \hat{a}_{1}^{\dag }\hat{a}_{1}+\hat{a}_{2}^{\dag }%
\hat{a}_{2}\right] +i\kappa \left[ \hat{a}_{1}^{\dag }\hat{a}_{2}^{\dag }-%
\hat{a}_{1}\hat{a}_{2}\right] .
\end{equation}%
The dynamical equation can be obtained from the Heisenberg equation (we set $%
\hbar=1$)
\begin{eqnarray}
i\partial _{t}\hat{a}_{j} &=&[\hat{a}_{j},\mathcal{H}]  \nonumber \\
i\partial _{t}\hat{a}_{j}^{\dag } &=&[\hat{a}_{j}^{\dag },\mathcal{H}].
\end{eqnarray}%
Using the commutation relations $[\hat{a}_{j},\hat{a}_{j^{\prime }}^{\dag
}]=\delta _{j,j^{\prime }}$ and $[\hat{a}_{j},\hat{a}_{j^{\prime }}]=0$, it
is straightforward to show that
\begin{eqnarray}
i\partial _{t}\hat{a}_{1} &=&\delta \hat{a}_{1}+i\kappa \hat{a}_{2}^{\dag }
\nonumber \\
i\partial _{t}\hat{a}_{2}^{\dag } &=&i\kappa \hat{a}_{1}-\delta \hat{a}%
_{2}^{\dag }.
\end{eqnarray}%
Thus we obtain the dynamical equation
\begin{equation}
i\partial _{t}\left(
\begin{array}{cc}
\hat{a}_{1}, & \hat{a}_{2}^{\dag }%
\end{array}%
\right) ^{T}=H_{APT}\left(
\begin{array}{cc}
\hat{a}_{1}, & \hat{a}_{2}^{\dag }%
\end{array}%
\right) ^{T},
\end{equation}%
with the non-Hermitian dynamical Hamiltonian matrix given by
\begin{equation}
H_{APT}=\delta \sigma _{z}+i\kappa \sigma _{x}=\left(
\begin{array}{cc}
\delta & i\kappa \\
i\kappa & -\delta%
\end{array}%
\right) .  \label{Eq:HAPTS5}
\end{equation}%
It is easy to show that the dynamical Hamiltonian possesses $\mathcal{APT}$%
-symmetry with the anticommutation relation $\mathcal{PT}H_{APT}=-H_{APT}%
\mathcal{PT}$, where parity operator $\mathcal{P}=\sigma _{x}$ and
time-reversal operator $\mathcal{T}=\mathcal{K}$ the complex conjugate. We
want to point out that the $\mathcal{APT}$-symmetric Hamiltonian can be
written as $H_{APT}=iH_{PT}$ with $H_{PT}=-i\delta \sigma _{z}+\kappa \sigma
_{x}$ satisfying the $\mathcal{PT}$-symmetry $\mathcal{PT}H_{PT}=H_{PT}%
\mathcal{PT}$.

The above discussion is gauge independent. This can be shown by considering
a different gauge choice, where we replace $\hat{a}_{2}$ by $i\hat{a}_{2}$
(i.e., $\hat{a}_{2}^{\dag }$ by $-i\hat{a}_{2}^{\dag }$). The
second-quantized Hamiltonian now reads
\begin{equation}
\mathcal{H}=\delta \left[ \hat{a}_{1}^{\dag }\hat{a}_{1}+\hat{a}_{2}^{\dag }%
\hat{a}_{2}\right] +\kappa \left[ \hat{a}_{1}^{\dag }\hat{a}_{2}^{\dag }+%
\hat{a}_{1}\hat{a}_{2}\right] .
\end{equation}%
Using the Heisenberg equation $i\partial _{t}\hat{a}_{j}=[\hat{a}_{j},%
\mathcal{H}]$ and commutation relations, we obtain%
\begin{eqnarray}
i\partial _{t}\hat{a}_{1} &=&\delta \hat{a}_{1}+\kappa \hat{a}_{2}^{\dag }
\nonumber \\
i\partial _{t}\hat{a}_{2}^{\dag } &=&-\kappa \hat{a}_{1}-\delta \hat{a}%
_{2}^{\dag }.
\end{eqnarray}%
The dynamical Hamiltonian in $i\partial _{t}(\hat{a}_{1},\hat{a}_{2}^{\dag
})^{T}=H_{APT}(\hat{a}_{1},\hat{a}_{2}^{\dag })^{T}$ becomes
\begin{equation}
H_{APT}=\delta \sigma _{z}+i\kappa \sigma _{y}=\left(
\begin{array}{cc}
\delta & \kappa \\
-\kappa & -\delta%
\end{array}%
\right) ,
\end{equation}%
which also possesses the $\mathcal{APT}$-symmetry $\mathcal{PT}%
H_{APT}=-H_{APT}\mathcal{PT}$, with $\mathcal{T}H_{APT}=H_{APT}\mathcal{T}$
and $\mathcal{P}H_{APT}=-H_{APT}\mathcal{P}$. We can again write the
Hamiltonian as $H_{APT}=iH_{PT}$ with $H_{PT}=-i\delta \sigma _{z}+\kappa
\sigma _{y}$ satisfying the $\mathcal{PT}$-symmetry $\mathcal{PT}%
H_{PT}=H_{PT}\mathcal{PT}$, with $\mathcal{T}H_{PT}=-H_{PT}\mathcal{T}$ and $%
\mathcal{P}H_{PT}=-H_{PT}\mathcal{P}$.

\subsection{S2. $\mathcal{APT}$ symmetry, EP and Two-mode squeezing}

Consider a general non-Hermitian Hamiltonian matrix $H$, which satisfies the
$\mathcal{APT}$ symmetry $\{\mathcal{PT},H\}=0$. We define $|R\rangle $ to
be the eigenstate of $H$ with $H|R\rangle =E|R\rangle $ and $E$ the
eigenvalue. We thus have $H\mathcal{PT}|R\rangle =-\mathcal{PT}H|R\rangle
=-E^{\ast }\mathcal{PT}|R\rangle $. In the $\mathcal{APT}$-symmetric phase,
the state $|R\rangle $ possesses $\mathcal{PT}$ symmetry $\mathcal{PT}%
|R\rangle =|R\rangle $, and we have $H\mathcal{PT}|R\rangle =H|R\rangle
=H|R\rangle $. Therefore, we have $E=-E^{\ast }$ in the $\mathcal{APT}$%
-symmetric phase, which must be purely imaginary. In the $\mathcal{APT}$%
-broken phase, the two eigenstates are different $\mathcal{PT}|R\rangle \neq
|R\rangle $, and we have a pair of eigenvalues $(E,-E^{\ast })$, which are
purely real up to a constant shift. Similar to the $\mathcal{PT}$ symmetry,
the $\mathcal{APT}$ symmetry breaking point also gives the EP which
corresponds to the change of eigenenergies from purely imaginary to purely
real.

The $\mathcal{APT}$-symmetric matrix $H_{APT}$ in Eq.~\ref{Eq:HAPTS5} is
non-Hermitian, thus has biorthogonal left and right eigenstates, $%
H_{APT}|R_{\pm }\rangle =\lambda _{\pm }|R_{\pm }\rangle $ and $%
H_{APT}^{\dag }|L_{\pm }\rangle =\lambda _{\pm }^{\ast }|L_{\pm }\rangle $.
In the pseudo-$\mathcal{APT}$-broken region $|\kappa |<|\delta |$, the
eigenvalues $\lambda _{\pm }=\pm \sqrt{\delta ^{2}-\kappa ^{2}}$, with
eigenstates $|R_{\pm }\rangle =\frac{i\kappa }{2\lambda _{\pm }}(\frac{%
\delta \pm \sqrt{\delta ^{2}-\kappa ^{2}}}{i\kappa },1)^{T}$, and $|L_{\pm
}\rangle =(1,\frac{\delta \mp \sqrt{\delta ^{2}-\kappa ^{2}}}{i\kappa })^{T}$%
. In the pseudo-$\mathcal{APT}$-symmetric region $|\kappa |>|\delta |$, $%
\lambda _{\pm }=\pm i\sqrt{\kappa ^{2}-\delta ^{2}}$ with $|R_{\pm }\rangle =%
\frac{i\kappa }{2\lambda _{\pm }}(\frac{\delta \pm i\sqrt{\kappa ^{2}-\delta
^{2}}}{i\kappa },1)^{T}$, and $|L_{\pm }\rangle =(1,\frac{\delta \pm i\sqrt{%
\kappa ^{2}-\delta ^{2}}}{i\kappa })^{T} $. We have two eigenmodes $\hat{b}%
_{+}\propto \langle L_{+}|\vec{a}\rangle $ and $\hat{b}_{-}^{\dag }\propto
\langle L_{-}|\vec{a}\rangle $ with $|\vec{a}\rangle =(\hat{a}_{1},\hat{a}%
_{2}^{\dag })^{T}$. We project the operators onto the eigenmodes, then the
field operators at time $t$ become
\begin{equation}
\left(
\begin{array}{c}
\hat{a}_{1}(t) \\
\hat{a}_{2}^{\dag }(t)%
\end{array}%
\right) =\left(
\begin{array}{cc}
A & B \\
B^{\ast } & A^{\ast }%
\end{array}%
\right) \left(
\begin{array}{c}
\hat{a}_{1}(0) \\
\hat{a}_{2}^{\dag }(0)%
\end{array}%
\right)
\end{equation}%
with the transfer matrix
\begin{equation}
\left(
\begin{array}{cc}
A & B \\
B^{\ast } & A^{\ast }%
\end{array}%
\right) =\sum_{s=\pm }|R_{s}\rangle e^{-i\lambda _{s}t}\langle L_{s}|,
\end{equation}%
where
\begin{eqnarray}
A &=&\sum_{s=\pm }\frac{(\lambda _{s}-\delta )e^{i\lambda _{s}t}}{2\lambda
_{s}} \\
B &=&\sum_{s=\pm }\frac{\kappa e^{i\lambda _{s}t}}{2i\lambda _{s}}.
\end{eqnarray}

At the EP $\kappa =\pm \delta $, the two eigenmodes coalesce to a single
mode $\sim \hat{a}_{1}\pm i\hat{a}_{2}^{\dag }$, which satisfies $i\partial
_{t}[\hat{a}_{1}(t)\pm i\hat{a}_{2}^{\dag }(t)]=2\delta \lbrack \hat{a}%
_{1}(0)\mp i\hat{a}_{2}^{\dag }(0)]$ and depends on its time-independent
orthogonal mode $\sim \hat{a}_{1}\mp i\hat{a}_{2}^{\dag }$. Therefore, we
have the solution $A=1-i\delta t$, $B=\kappa t$. We can define the
quadrature operators $\hat{X}_{j}(\varphi ,t)=[e^{-i\varphi }\hat{a}%
_{j}(t)+h.c.]/2$ and $\hat{P}_{j}(\varphi ,t)=[e^{-i\varphi }\hat{a}%
_{j}(t)-h.c.]/2i$. It can be shown that $\hat{X}_{1}(\varphi _{+},t)\pm \hat{%
X}_{2}(\varphi _{+},t)=S^{\pm 1}[\hat{X}_{1}(\varphi _{-},0)\pm \hat{X}%
_{2}(\varphi _{-},0)]$ and $\hat{P}_{1}(\varphi _{+},t)\mp \hat{P}%
_{2}(\varphi _{+},t)=S^{\pm 1}[\hat{P}_{1}(\varphi _{-},0)\mp \hat{P}%
_{2}(\varphi _{-},0)]$, with the squeezing factor $S=A_{0}+B_{0}\geq 1$
(with $A_{0}=|A|,B_{0}=|B|$) and the angles $\varphi _{\pm }=(\mathrm{Arg}%
[B]\pm \mathrm{Arg}[A])/2$. The oscillation behavior of $S$ (as well as $A$
and $B$) in the pseudo-$\mathcal{APT}$-broken region is due to the
interference between two eigenmodes.

It can be easily shown that the solutions $\hat{a}_{1}(t),\hat{a}_{2}(t)$
satisfy the Bosonic commutation relation. Since the corresponding
second-quantized Hamiltonian $\mathcal{H}=\delta \left[ \hat{a}_{1}^{\dag }%
\hat{a}_{1}+\hat{a}_{2}^{\dag }\hat{a}_{2}\right] +i\kappa \left[ \hat{a}%
_{1}^{\dag }\hat{a}_{2}^{\dag }-\hat{a}_{1}\hat{a}_{2}\right] $ is
Hermitian, we denote that our system possesses a pseudo-$\mathcal{APT}$
symmetry. There are no Langevin noises in our model.

For comparison, we briefly discuss the $\mathcal{PT}$-symmetry physics in
previous studies which generally utilize the control of gain/loss, whose
dynamical equation is
\begin{equation}
i\partial _{t}\left(
\begin{array}{c}
\hat{a}_{1} \\
\hat{a}_{2}%
\end{array}%
\right) =\left(
\begin{array}{cc}
i\delta & \kappa \\
\kappa & -i\delta%
\end{array}%
\right) \left(
\begin{array}{c}
\hat{a}_{1} \\
\hat{a}_{2}%
\end{array}%
\right) +\left(
\begin{array}{c}
\hat{f}_{1} \\
\hat{f}_{2}%
\end{array}%
\right) ,  \label{eq:bdg2SS}
\end{equation}%
where $\hat{f}_{1}$ and $\hat{f}_{2}$ are Langevin noise operators. To study
the non-Hermitian physics in quantum realm, we have replaced the classic
wave amplitudes by the quantum field operators. In the classical domain,
these Langevin noises are removed because their zero expectation values.
However, this will lead to a problem in the quantum domain where the field
operator solutions without Langevin noise operators do not obey the Bosonic
commutation relations. Therefore Langevin noise operators must be included
in the above dynamical equation, which breaks the $\mathcal{PT}$ symmetry.

We have focused on the dynamics of the system in the main text. To see how
the quantum pseudo-$\mathcal{APT}$ symmetry is related with the ground state
of the system, we consider the second-quantized Hamiltonian $\mathcal{H}%
=\delta \left[ \hat{a}_{1}^{\dag }\hat{a}_{1}+\hat{a}_{2}^{\dag }\hat{a}_{2}%
\right] +i\kappa \left[ \hat{a}_{1}^{\dag }\hat{a}_{2}^{\dag }-\hat{a}_{1}%
\hat{a}_{2}\right] $. In the pseudo-$\mathcal{APT}$-broken region, the two
eigenmodes read $\hat{b}_{+}=\frac{\kappa }{\sqrt{2\lambda _{0}\delta
-2\lambda _{0}^{2}}}\langle L_{+}|\vec{a}\rangle $ and $\hat{b}_{-}^{\dag }=%
\frac{\kappa }{\sqrt{2\lambda _{0}\delta +2\lambda _{0}^{2}}}\langle L_{-}|%
\vec{a}\rangle $. We have $\mathcal{H}=\lambda _{0}(\hat{b}_{+}^{\dag }\hat{b%
}_{+}+\hat{b}_{-}^{\dag }\hat{b}_{-})$ and $\hat{b}_{\pm }$ satisfy the
bosonic commutation relation. The ground state satisfy $\hat{b}_{\pm }|\text{%
GS}\rangle =0$, which is a two-mode squeezed state in the $\hat{a}_{1,2}$
basis, \textit{i.e.}, $|\text{GS}\rangle =\sum_{n}\sqrt{1-|q|^{2}}%
q^{n}|n,n\rangle $ with $q=i\frac{\delta -\lambda _{0}}{\kappa }$ and $%
|n,n\rangle $ the Fock state in $\hat{a}_{1,2}$ basis. In the pseudo-$%
\mathcal{APT}$-symmetric region, the two eigenmodes no longer satisfy
bosonic commutation relation, therefore the second-quantized Hamiltonian
cannot be diagonalized and has no well defined ground state. The system is
always dynamically unstable and the energy of the system is not bounded from
below. We want to emphasize that, here we are interested in the dynamics of
the system rather than its ground state, therefore, the system is not
necessarily stable or with energy bounded from below. For a physical process
such as the parametric down conversion, the dynamics are well characterized
by our model. In a realistic experiment, the evolution time and thereby the
photon number are always finite, and the dynamical model is physical.

Recall that the squeezing factor $S$ is defined based on the dynamical
evolution of the field operators (i.e., $\hat{X}_{1}(\varphi _{+},t)\pm \hat{%
X}_{2}(\varphi _{+},t)=S^{\pm 1}[\hat{X}_{1}(\varphi _{-},0)\pm \hat{X}%
_{2}(\varphi _{-},0)]$), therefore, $S$ is independent of the initial state.
However, the variance of $\hat{X}_{1}(\varphi ,t)\pm \hat{X}_{2}(\varphi ,t)$
does depend on the initial state. If we start from the state $|\text{GS}%
\rangle $, then the variance of $\hat{X}_{1}(\varphi ,t)\pm \hat{X}%
_{2}(\varphi ,t)$ should not change with time for all $\varphi $, since $|%
\text{GS}\rangle $ is the ground state of $\mathcal{H}$ and thus $|\text{GS}%
(t)\rangle =|\text{GS}(t=0)\rangle $. On the other hand, the dynamics of the
field operator in Heisenberg picture give $\hat{X}_{1}(\varphi _{+},t)\pm
\hat{X}_{2}(\varphi _{+},t)=S^{\pm 1}[\hat{X}_{1}(\varphi _{-},0)\pm \hat{X}%
_{2}(\varphi _{-},0)]$, which indicate that the two-mode variance along $%
\varphi _{+}$ at time $t$ is relatively squeezed with respect to the
two-mode variance along $\varphi _{-}$ at time $0$. This means that the
two-mode variance along $\varphi _{+}$ is $S^{\pm 2}$ times of the two-mode
variance along $\varphi _{-}$ for an initial state $|\text{GS}\rangle $
whose two-mode variance is time-independent. It can be shown that $|\text{GS}%
\rangle $, a two-mode squeezed state, does satisfy the above relation.

\subsection{S3. Quantum Fisher information}

We focus on the initial coherent state $|\psi _{0}\rangle =|\alpha
_{1},\alpha _{2}\rangle $, and the final state reads $|\psi (t)\rangle =e^{-i%
\mathcal{H}t}|\psi _{0}\rangle $ with the second-quantized Hamiltonian $%
\mathcal{H}=\delta \left[ \hat{a}_{1}^{\dag }\hat{a}_{1}+\hat{a}_{2}^{\dag }%
\hat{a}_{2}\right] +i\kappa \left[ \hat{a}_{1}^{\dag }\hat{a}_{2}^{\dag }-%
\hat{a}_{1}\hat{a}_{2}\right] $. We write $|\psi _{0}\rangle =\hat{D}%
_{1}(\alpha _{1})\hat{D}_{2}(\alpha _{2})|0,0\rangle $ with displacement
operators $\hat{D}_{j}(\alpha _{j})=e^{\alpha _{j}\hat{a}_{j}^{\dag }-h.c.}$
and vacuum state $|0,0\rangle $. Now we are working in the Schr\"{o}dinger
picture and the time-independent field operators are $\hat{a}_{j}=\hat{a}%
_{j}(0)$. Then we have $|\psi (t)\rangle =\hat{D}_{1}(\alpha _{1}^{\prime })%
\hat{D}_{2}(\alpha _{2}^{\prime })\sum_{n}\sqrt{1-|q|^{2}}q^{n}|n,n\rangle $%
, with $\alpha _{j}^{\prime }=A\alpha _{j}+B\alpha _{\bar{j}}$, $q=\frac{B}{%
A^{\ast }}$ and $|n,n\rangle $ the Fock state. In particular, $|\psi
(t)\rangle =[e^{-i\mathcal{H}t}\hat{D}_{1}(\alpha _{1})\hat{D}_{2}(\alpha
_{2})e^{i\mathcal{H}t}]e^{-i\mathcal{H}t}|0,0\rangle $. Using $e^{-i\mathcal{%
H}t}\hat{a}_{j}e^{i\mathcal{H}t}=A(-t)\hat{a}_{j}+B(-t)\hat{a}_{\bar{j}%
}^{\dag }$, $A(-t)=A^{\ast }(t)$ and $B(-t)=-B(t)$, we obtain $[e^{-i%
\mathcal{H}t}\hat{D}_{1}(\alpha _{1})\hat{D}_{2}(\alpha _{2})e^{i\mathcal{H}%
t}]=\hat{D}_{1}(\alpha _{1}^{\prime })\hat{D}_{2}(\alpha _{2}^{\prime })$.
From $\hat{a}_{j}|0,0\rangle =0$, we have $[e^{-i\mathcal{H}t}\hat{a}_{j}e^{i%
\mathcal{H}t}]e^{-i\mathcal{H}t}|0,0\rangle =[A^{\ast }\hat{a}_{j}-B\hat{a}_{%
\bar{j}}]e^{-i\mathcal{H}t}|0,0\rangle =0$, thus $e^{-i\mathcal{H}%
t}|0,0\rangle =\sqrt{1-|q|^{2}}\sum_{n}q^{n}|n,n\rangle $, which is a
two-mode squeezed vacuum state. The quantum Fisher information is calculated
as $F_{\kappa }=4\langle \partial _{\kappa }\psi (t)|\partial _{\kappa }\psi
(t)\rangle -4|\langle \partial _{\kappa }\psi (t)|\psi (t)\rangle |^{2}$~%
\cite{RevModPhys.90.035005S}. We can obtain the expression given by Eq. (7)
in the main text after some algebraic manipulations.

For the sensing in the $\mathcal{APT}$-broken (stable) region, the final
state has no squeezing (\textit{i.e.}, $S=1$) at the working point $t=nT$,
nevertheless, the QFI possesses the divergent scaling $F_{\kappa }\sim
(2n\pi)^2 \lambda _{0}^{-6}$ in the vicinity of the EP with $\lambda_0\sim0$
and $\lambda_0t=n\pi$. Even we consider the squeezing during the evolution,
which reaches its maximum $S_m\sim \lambda_0^{-1}$ at time $t=nT+T/2$, we
have $F_{\kappa }\sim(2n\pi)^2 S_{m}^{6}$. Therefore, QFI can be very large
without strong squeezing during the whole evolution (especially for a large $%
n$).
Also, for our EP-based sensing, the sensing precision of our measurement
scheme is at the same order as quantum Cram\'er-Rao bound set by the QFI.
Our scheme works in a wide range of parameter space on the stable side of
the EP. Moreover, the sensitivity as a function of evolution time reaches
its peak values at the working points (i.e., $t=nT$). These peaks have a
large width which roughly equals to 1/4 the peak separation, thus high
sensitivity can still be achieved even when the evolution time is slightly
away from the working point.

In comparison, for the $\mathcal{APT}$-symmetric (unstable) region, we find $%
F_{\kappa }\sim e^{4\lambda _{0}t}\lambda _{0}^{-6}$ when $e^{\lambda _{0}t}$
is much larger than $e^{-\lambda _{0}t}$. In the vicinity of the EP with $%
\lambda _{0}\sim 0$, the QFI possesses the divergent scaling $F_{\kappa
}\sim \lambda _{0}^{-6}$ for a fixed $\lambda _{0}t$ (e.g., an evolution
time $\lambda _{0}t=n\pi $), which is similar to the stable region. Away
from the EP with $\lambda _{0}\sim 1$, the QFI $F_{\kappa }\sim e^{4t}$
increases exponentially and can be very high after a long time. However, in
the unstable region, the high QFI originates mainly from the squeezing
instead of the EP, since the final state has a large squeezing factor $S\sim
e^{\lambda _{0}t}\lambda _{0}^{-1}$. In the vicinity of the EP with $\lambda
_{0}\sim 0$ and $\lambda _{0}t=n\pi $, we have $F_{\kappa }\sim \lambda
_{0}^{-2}S^{4}\sim e^{-2n\pi }S^{6}$. Thus away from the EP with $\lambda
_{0}\sim 1$, we have $F_{\kappa }\sim S^{4}$. High QFI in the unstable
region requires much stronger squeezing than the stable region (especially
for a large $n$), which is generally hard to achieve in experiment since
large numbers of excited particles $\sim S^{2}$ would be generated during
the sensing process. On the unstable side of the EP, a larger $\lambda _{0}$
would lead to a larger QFI for a fixed evolution time, for our measurement
scheme with $\lambda _{0}$-dependent evolution time $t=n\pi /\lambda _{0}$,
both the susceptibility and variance of the measured quadrature are very
large, leading to a sensitivity $\Delta _{\kappa }^{-2}\sim \lambda
_{0}^{-4} $ near the EP. This sensitivity is worse compared with the stable
side. Therefore, for sensors working on the unstable side, new novel
measurement scheme is required to reach sensitivity of the order of QFI
(i.e., quantum Cram\'{e}r-Rao bound). We want to emphasize that, it might be
possible to operate within the unstable region for nonlinear optics, but for
the BEC, it is challenging since large numbers of excitations are generated
during the sensing process. Our sensing scheme, which works very well in the
stable side, can be implemented in both nonlinear optics and BEC.

\begin{figure}[t]
\includegraphics[width=0.8\linewidth]{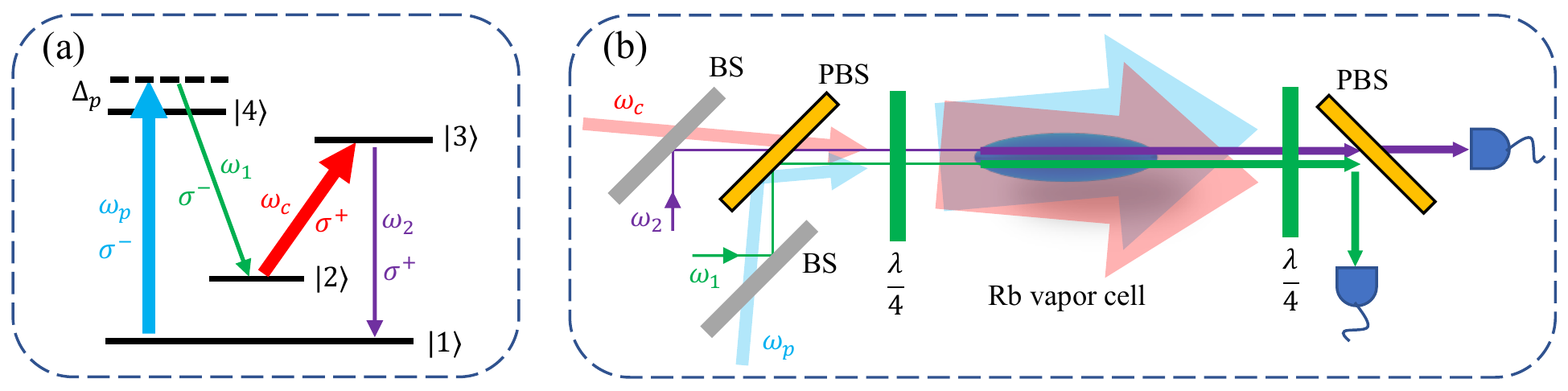}
\caption{Experimental realization based on nonlinear quantum optics. (a) $%
^{85}$Rb atomic energy-level diagram: $|1\rangle =|5S_{1/2},F=2\rangle $, $%
|2\rangle =|5S_{1/2},F=3\rangle $, $|3\rangle =|5P_{1/2},F=3\rangle $, $%
|4\rangle =|5P_{3/2},F=3\rangle $. (b) Schematic of the experimental setup
and geometrical arrangement of four interacting fields. BS and PBS represent
the beam splitter and polarized beam splitter.}
\label{fig:S1}
\end{figure}

\subsection{S4. Experimental consideration}

\emph{{\color{blue}Nonlinear quantum optics}.---}Our model can be realized
by multi-wave mixing processes in nonlinear optical medium. Here we give
more details about the experimental scheme based on four-wave mixing (the
scheme based on three-wave mixing SPDC is similar). We consider the same
setup as in~\cite{PhysRevLett.123.193604S}, with $^{85}$Rb atomic ensemble
(prepared on the ground state $|1\rangle $) as the nonlinear medium. The
atomic energy-level diagram, geometrical arrangement of four interacting
fields and schematic of the experimental setup are shown in Fig.~\ref{fig:S1}%
. A pump laser ($\omega _{p}$) is blue detuned by $\Delta _{p}$ from the
atomic transition $|1\rangle \rightarrow |4\rangle $ and a weak Stokes field
($\omega _{1}$) follows $|4\rangle \rightarrow |2\rangle $. Another strong
coupling laser ($\omega _{c}$) is on resonance at $|2\rangle \rightarrow
|3\rangle $ and the weak anti-Stokes field ($\omega _{2}$) drives the
transition $|3\rangle \rightarrow |1\rangle $. We consider resonant mixing
with energy conservation $\omega _{p}+\omega _{c}=\omega _{1}+\omega _{2}$.

The coupled equations are
\begin{equation}
i\partial _{z}\left(
\begin{array}{c}
\hat{a}_{1} \\
\hat{a}_{2}^{\dag }%
\end{array}%
\right) =\left(
\begin{array}{cc}
-\Delta k/2 & -\tilde{\kappa} \\
\tilde{\kappa} & \Delta k/2%
\end{array}%
\right) \left(
\begin{array}{c}
\hat{a}_{1} \\
\hat{a}_{2}^{\dag }%
\end{array}%
\right) .  \label{eq:FWM}
\end{equation}%
Here $\Delta k=(\mathbf{k}_{1}+\mathbf{k}_{2}-\mathbf{k}_{p}-\mathbf{k}%
_{c})\cdot \hat{\mathbf{z}}=k_{1}+k_{2}-(k_{c}+k_{p})\cos (\theta _{cp})$ is
the phase mismatch with $k_{j}=|\mathbf{k}_{j}|$ the corresponding wave
vector, $\hat{\mathbf{z}}$ the unit vector along propagation direction $z$
and $\theta _{cp}$ the relative angle between the pumping and coupling
beams. $\Delta k$ can be tuned by slightly changing the propagation
direction of the pumping or coupling beam. Typically, the wavelength of
pumping (coupling) laser is $780$ nm ($795$ nm), and $\theta
_{cp}=0.2^{\circ }$, so we have $\Delta k\simeq 100$ rad/m. $\tilde{\kappa}$
denotes the real nonlinear coupling coefficient~\cite%
{PhysRevLett.123.193604S}
\begin{equation}
\tilde{\kappa}=\frac{N_{a}\sqrt{\sigma _{13}\sigma _{24}\gamma _{13}\gamma
_{14}}}{|\Omega _{c}|^{2}+4\gamma _{13}\gamma _{12}}\frac{\Omega _{p}\Omega
_{c}}{2\Delta _{p}},
\end{equation}%
where $\Omega _{c}$ and $\Omega _{p}$ are coupling and pump Rabi
frequencies, $\gamma _{ij}$ is the dephasing or decay rate between the
states $|i\rangle $ and $|j\rangle $, $N_{a}$ is the atomic density, $\sigma
_{ij}$ is the absorption cross section of the atomic transition $|i\rangle
\rightarrow |j\rangle $. Typically, the detuning $\Delta _{p}$ is of the
order of $100$ MHz. $\tilde{\kappa}$ can be tuned by $\Omega _{c}$, $\Omega
_{p}$ and $\Delta _{p}$. $\Delta _{p}$ is tuned by changing the frequency of
$\omega _{p}$ with the change below 100 MHz, which is enough to
significantly tune $\tilde{\kappa}$. Notice that $\Delta k\propto
k_{p}\propto \omega _{p}$, thus 100 MHz frequency change in $\omega _{p}$
only slightly modifies $\Delta k$ by $\frac{100\text{MHz}}{\omega _{p}}\cdot
\Delta k\sim 10^{-5}$ rad/m, which is negligible.

With a proper gauge choice $\hat{a}_{2}^{\dag }\rightarrow \hat{a}_{2}^{\dag
}e^{i\pi /2}$, we obtain
\begin{equation}
i\partial _{z}\left(
\begin{array}{c}
\hat{a}_{1} \\
\hat{a}_{2}^{\dag }%
\end{array}%
\right) =\left(
\begin{array}{cc}
\delta & i{\kappa } \\
i{\kappa } & -\delta%
\end{array}%
\right) \left(
\begin{array}{c}
\hat{a}_{1} \\
\hat{a}_{2}^{\dag }%
\end{array}%
\right) =H\left(
\begin{array}{c}
\hat{a}_{1} \\
\hat{a}_{2}^{\dag }%
\end{array}%
\right) ,  \label{eq:FWM2}
\end{equation}%
with $\delta =-\Delta k/2$ and $\kappa =-\tilde{\kappa}$.

For the quantum sensing, we can tune $\Delta k$ to the work points
satisfying $\lambda _{0}L=n\pi $ with $z=L$ the length of the nonlinear
medium. The initial state can be obtained by injecting weak input coherent
fields in $\hat{a}_{j}$ (see Fig.~\ref{fig:S1}), and the measurement is
realized by standard homodyne detection. It can be shown that the
second-quantized Hamiltonian of Eq.~\ref{eq:FWM} is $\mathcal{H}=-\delta %
\left[ \hat{a}_{1}^{\dag }\hat{a}_{1}+\hat{a}_{2}^{\dag }\hat{a}_{2}\right]
-\kappa \left[ \hat{a}_{1}^{\dag }\hat{a}_{2}^{\dag }+\hat{a}_{1}\hat{a}_{2}%
\right] $, which is equivalent to our previous second-quantized Hamiltonian
up to a gauge transformation.

\begin{figure}[t]
\includegraphics[width=0.8\linewidth]{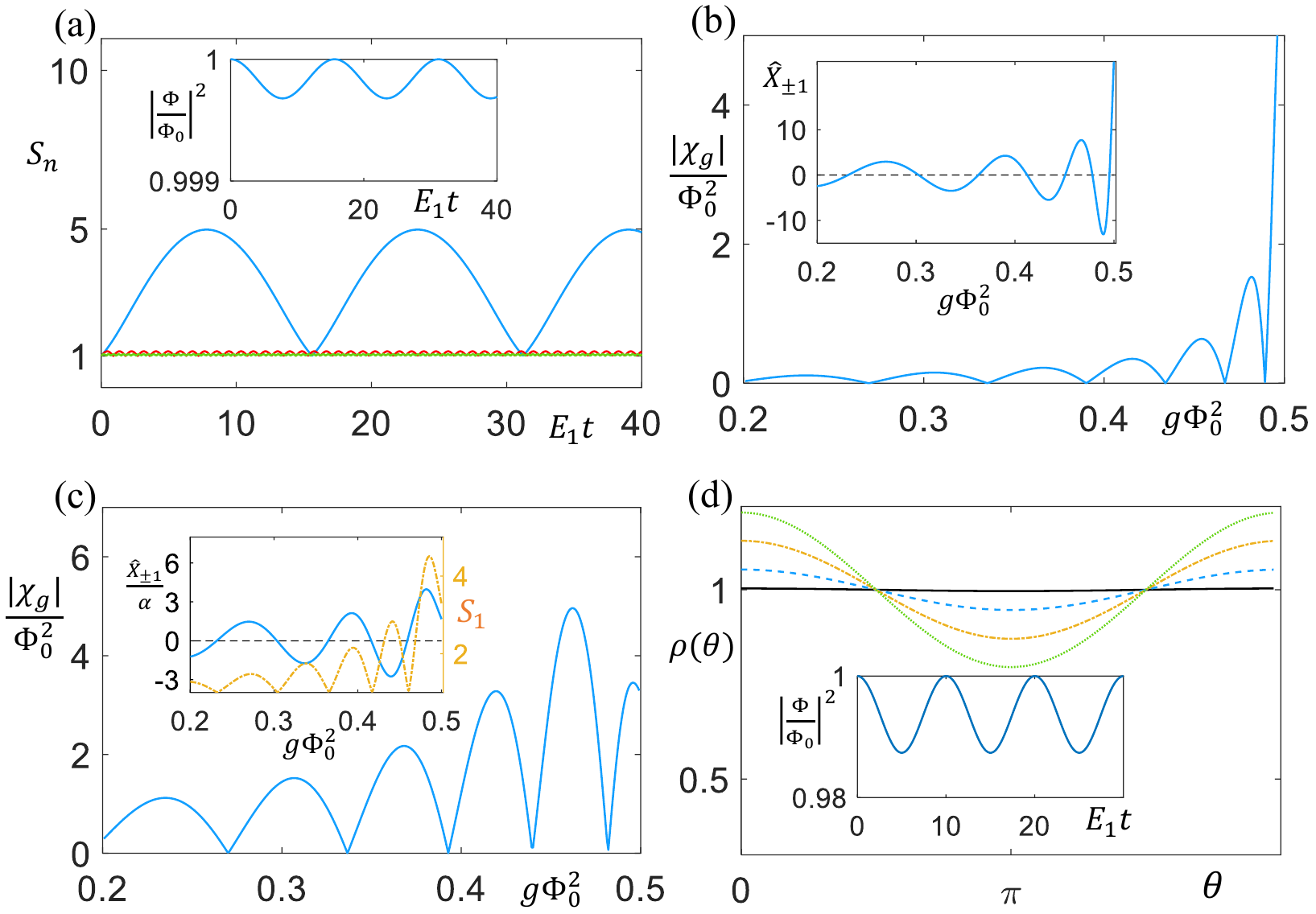}
\caption{Quantum sensing based on atomic BEC trapped in a ring. (a) The
squeezing factors $S_{n}$ as functions of evolution time for fields $\hat{%
\protect\psi}_{\pm n}$ with $g\Phi _{0}^{2}=0.48E_{1}$ and $\protect\alpha %
=2 $. A larger $n$ has a weaker squeezing factor. $n=1,2,3$ are shown. Inset
shows how $\Phi (t)$ changes with time. (b) and (c) The susceptibility (in
unit of $10^{3}$) as a function of $g\Phi _{0}^{2}$ at time $E_{1}t=30$ for $%
\protect\alpha =2$ and $\protect\alpha =20$. Inset of (b) shows the
corresponding quadrature. Inset of (c) shows the corresponding quadrature
(solid line) and squeezing factor (dash-dotted line). (d) Density modulation
along azimuthal angle $\protect\theta $ for $E_{1}t=30$ and $\protect\alpha %
=20$. The solid, dashed, dash-dotted and dotted lines correspond to $g\Phi
_{0}^{2}=0.46E_{1},0.462E_{1},0.465E_{1},0.468E_{1}$, respectively, and the
corresponding quadratures and visibilities are $\langle \hat{X}_{\pm
1}\rangle =4.41,14.3,29.1,43.2$ and $v=0.3\%,5.3\%,12.9\%,20.4\%$. Inset
shows the condensed atom number during the evolution for $g\Phi
_{0}^{2}=0.46E_{1}$ and $\protect\alpha =20$.}
\label{fig:S2}
\end{figure}

\emph{{\color{blue}Ultra-cold atoms}.---}We can substitute the expansion of
the BEC field operator
\begin{equation}
\hat{\Psi}(\theta ,t)=e^{-i\mu t-i\pi /4}\Phi (t)+e^{-i\mu t}\sum_{n\neq 0}%
\hat{\psi}_{n}(t)\frac{e^{in\theta }}{\sqrt{2\pi }},  \label{eq:fieldS}
\end{equation}%
into the nonlinear Schr\"{o}dinger equation
\begin{equation}
i\hbar \partial _{t}\hat{\Psi}(\theta ,t)=[\frac{\hat{p}^{2}}{2m}-{g}\hat{%
\Psi}(\theta ,t)^{\dag }\hat{\Psi}(\theta ,t)]\hat{\Psi}(\theta ,t),
\label{eq:nlseS}
\end{equation}%
which gives the Hartree-Fock Bogoliubov equation
\begin{eqnarray}
i\partial _{t}{\Phi } &=&-i\frac{g}{2\pi }\sum_{n}\langle \hat{\psi}_{n}\hat{%
\psi}_{-n}\rangle \Phi ^{\ast }-\frac{g}{2\pi }\sum_{n}\langle \hat{\psi}%
_{n}^{\dag }\hat{\psi}_{n}\rangle \Phi ,  \label{eq:HFbdg1} \\
i\partial _{t}\left(
\begin{array}{c}
\hat{\psi}_{n} \\
\hat{\psi}_{-n}^{\dag }%
\end{array}%
\right) &=&\left(
\begin{array}{cc}
n^{2}E_{1}-g|\Phi |^{2} & ig\Phi ^{2} \\
ig(\Phi ^{\ast })^{2} & -(n^{2}E_{1}-g|\Phi |^{2})%
\end{array}%
\right) \left(
\begin{array}{c}
\hat{\psi}_{n} \\
\hat{\psi}_{-n}^{\dag }%
\end{array}%
\right)  \label{eq:HFbdg2}
\end{eqnarray}%
In the stable pseudo-$\mathcal{APT}$-broken region, excitations are weak and
$\sum_{n}\langle \hat{\psi}_{n}\hat{\psi}_{-n}\rangle $ and $\sum_{n}\langle
\hat{\psi}_{n}^{\dag }\hat{\psi}_{n}\rangle $ are much smaller than $\Phi
^{2}$, therefore $\Phi $ is hardly affected by the excitations.

For the quantum sensing, we can first prepare the BEC to its ground state
with $g\simeq 0$. The initial coherent state for $n=\pm 1$ can be generated
by the Raman process with Laguerre-Gaussian beams carrying $\pm 1$ orbital
angular momentum. Then we increase $g$ to the working point near the EP (%
\textit{i.e.}, $2g|\Phi |^{2}=E_{1}$). For example, we can tune $g$ using
Feshbach resonance, which in turn can be used to measure the magnetic field
with ultra-high precision. The sensor operates in the pseudo-$\mathcal{APT}$%
-broken region which ensures the dynamical stability of the BEC. The
squeezing and excitation number for $n>1$ are negligible since $\delta $
increases quadratically with $n$. Moreover, the excitation number for $n=1$
is generally much smaller than the number of BEC atoms, so that $\Phi $ is
hardly affected during the evolution, which is confirmed by our numerical
simulation. The performance of the sensor (Fig.~\ref{fig:S2}) is very
similar as that shown in Fig.~2 in the main text. For the measurement
scheme, one can detect the density $\langle \hat{\Psi}^{\dag }(\theta ,t)%
\hat{\Psi}(\theta ,t)\rangle \simeq \Phi ^{2}+\frac{2\Phi }{\sqrt{2\pi }}%
\sum_{n\neq 0}\langle \hat{X}_{n}(\varphi ,t)\cos (n\theta )-\hat{P}%
_{n}(\varphi ,t)\sin (n\theta )\rangle $ with $\varphi =-\pi /4$ and $\hat{X}%
_{n}$, $\hat{P}_{n}$ the corresponding quadrature operators of $\hat{\psi}%
_{n}$. Notice that only $\langle \hat{X}_{\pm 1}(\varphi ,t)\rangle $ and $%
\langle \hat{P}_{\pm 1}(\varphi ,t)\rangle $ are nonzero (since the initial
excitations are prepared on $n=\pm 1$) and they show divergent
susceptibility with respect to $g$ or $E_{1}$.

In the following, we show the numerical simulations for the BEC-based
quantum sensing. We find that ultra-high precision can be still achieved
even if we consider weak back action of the excitations on the condensate
wave function, and the long-time dynamics would still be very sensitive to
system parameters around the pseudo-$\mathcal{APT}$ transition. In Fig.~\ref%
{fig:S2}(a), we show the squeezing factors and $|\Phi |^{2}$ as a function
of time by solving Eq.~\ref{eq:HFbdg1} and Eq.~\ref{eq:HFbdg2} numerically.
We see $|\Phi |^{2}$ is hardly affected even for parameters close to the
exceptional point. In Fig.~\ref{fig:S2}(b), we show how the quadratures $%
\langle \hat{X}_{\pm 1}(\varphi ,t)\rangle $ and the corresponding
susceptibilities change with $g|\Phi _{0}|^{2}$ (where $\Phi (0)=\Phi _{0}$)
for $\phi =-\pi /4$. In the simulation, we set $E_{1}=1$ as the unit, and
assume total atom number of the BEC is about $\Phi _{0}^{2}=10^{5}$, which
is typical in experiments. We also consider an initial excited state being
the coherent state with amplitudes $\alpha _{+1}=\alpha _{-1}=e^{i\pi
/4}\alpha $, then we have $\hat{X}_{+1}(-\pi /4,t)=\hat{X}_{-1}(-\pi /4,t)$,
and $\chi _{g}=\partial _{g}\hat{X}_{+1}$. In our simulation, we keep $n$ up
to $\pm 10$ and check that the results are hardly affected by increasing the
cutoff of $n$.

For our choice with $\alpha _{+1}=\alpha _{-1}=e^{i\pi /4}\alpha $, the
quadratures are $\langle \hat{X}_{\pm 1}(-\pi /4,0)\rangle =\mathrm{Re}%
[e^{i\pi /4}\alpha _{\pm 1}]=0$ and $\langle \hat{P}_{\pm 1}(-\pi
/4,0)\rangle =\mathrm{Im}[e^{i\pi /4}\alpha _{\pm 1}]=\alpha $. The initial
density reads
\begin{equation}
\langle \hat{\Psi}^{\dag }\hat{\Psi}\rangle \simeq \Phi _{0}^{2}+\frac{2\Phi
_{0}}{\sqrt{2\pi }}\sum_{n=\pm 1}\langle \hat{X}_{n}\rangle \cos (n\theta
)-\langle \hat{P}_{n}\rangle \sin (n\theta )=\Phi _{0}^{2}.
\end{equation}%
There is no density modulation initially. Notice that we have $\hat{P}_{+1}=%
\hat{P}_{-1}$ and $\hat{X}_{+1}=\hat{X}_{-1}$ for our choice of $\alpha
_{\pm 1}$, $\hat{X}_{\pm 1}$ and $\hat{P}_{\pm 1}$ correspond to density and
phase modulation of the wave function, and the total density is given by $%
\langle \hat{\Psi}^{\dag }\hat{\Psi}\rangle \simeq \Phi _{0}^{2}+\frac{4\Phi
_{0}}{\sqrt{2\pi }}\langle \hat{X}_{1}\rangle \cos (\theta )$. This can also
be seen from the initial total wave function which is $\langle \hat{\Psi}%
\rangle =e^{-i\pi /4}[\Phi _{0}+i\frac{2\alpha }{\sqrt{2\pi }}\cos (\theta
)] $. The initial excitation with nonzero $\hat{P}_{\pm 1}$ corresponds to
phase modulation. We can define the normalized density as
\begin{equation}
\rho (\theta ,t)=\frac{\langle \hat{\Psi}^{\dag }\hat{\Psi}\rangle }{\Phi
^{2}}\simeq 1+\frac{4}{\Phi _{0}\sqrt{2\pi }}\langle \hat{X}_{1}\rangle \cos
(\theta ),
\end{equation}%
therefore, the observable $\langle \hat{X}_{\pm 1}\rangle \simeq \frac{\Phi
_{0}\sqrt{2\pi }}{4}v$ can be detected by density imaging with $v=\frac{%
\mathrm{max}[\rho (\theta )]-\mathrm{min}[\rho (\theta )]}{\mathrm{max}[\rho
(\theta )]+\mathrm{min}[\rho (\theta )]}$ the visibility. Around the working
point, the system nearly returns to the initial state with $\langle \hat{X}%
_{\pm 1}\rangle \simeq 0$ and $\langle \hat{P}_{\pm 1}\rangle \simeq \alpha $%
, where $\langle \hat{X}_{\pm 1}\rangle $ and thereby the density modulation
is very sensitive to system parameters.

For a small $\alpha $ (e.g., $\alpha =2$), the excitation number and the
observable $\langle \hat{X}_{\pm 1}\rangle $ are small, therefore, the
density modulation is also weak, which makes the detection hard. To obtain
density modulation that is easy to detect, we need to use a larger $\alpha $%
. On the other hand, $\alpha $ should not be too large so that the
excitation number is much smaller than the ground-state atom number and the
system can still be characterized by the Hartree-Fock Bogoliubov equation.
There is a trade-off in choosing $\alpha $. We find that $\alpha =20$ can
lead to density modulation with visibility $\sim 20\%$, while keeping the
change in condensed atom number below $2\%$ (as shown in Figs.~\ref{fig:S2}c
and ~\ref{fig:S2}d). Such visibility could be observed using current imaging
technique, and the density modulation pattern is very sensitive to
parameters (e.g., the visibility changes from $0.3\%$ to $12.9\%$ as $g\Phi
_{0}^{2}$ changes from $0.46E_{1}$ to $0.465E_{1}$). The reason why weak
excitations can lead to strong modulation is that, $\rho (\theta )=1+v\cos
(\theta )$ corresponds to a wave function $\langle \hat{\Psi}\rangle \sim
\frac{1}{\sqrt{2\pi }}+\frac{v}{2\sqrt{2}}\frac{\cos (\theta )}{\sqrt{\pi }}$
(we ignore the phase modulation), and the excitation fraction [on excited
state $\frac{\cos (\theta )}{\sqrt{\pi }}$] is given by $(\frac{v}{2\sqrt{2}}%
)^{2}=v^{2}/8$. Therefore, even for a visibility $v\sim 20\%$, the
excitation fraction (or the relative change in $|\Phi (t)|^{2}$) is only
about $0.5\%$. The maximum excitation fraction is reached when the
quadrature $\langle \hat{X}_{1}\rangle $ and also the squeezing factor $%
S_{1} $ reach their maximum, which is away from the working point and the
corresponding susceptibility drops to 0. For $g\Phi _{0}^{2}=0.46E_{1}$ and $%
\alpha =20$, the maximum excitation fraction is about $1.3\%$ (see the inset
of Fig.~\ref{fig:S2}d at $E_{1}t\simeq 5,15,25,\cdots $), and the
corresponding maximum visibility is about $30\%$.

Notice that, if we start from the exceptional point $g\Phi_0^2=0.5E_1$, the
system will evolve back to the symmetry broken region due to the oscillation
of $\Phi^2(t)$ (the oscillation is stronger for larger $\alpha$), therefore,
the transition at $g\Phi_0^2=0.5E_1$ is no longer sharp, but behaves like a
crossover.

\begin{figure}[t]
\includegraphics[width=0.8\linewidth]{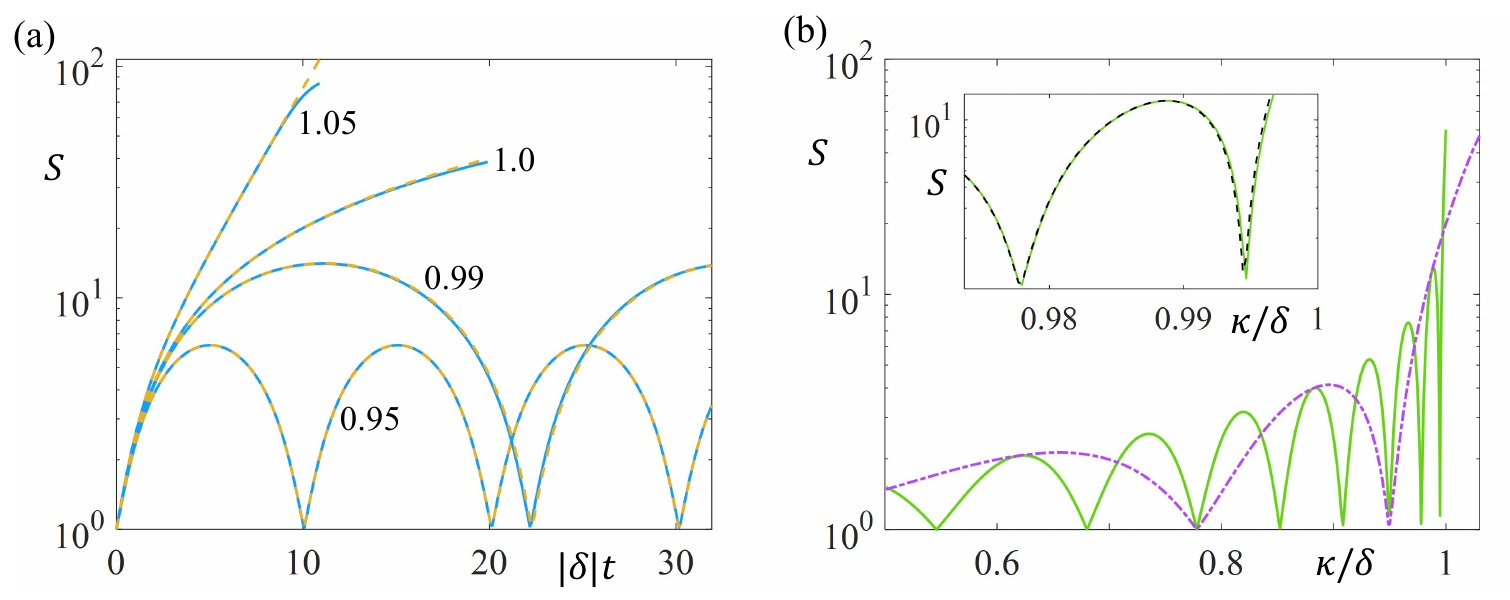}
\caption{Numerical simulations of the squeezing factor with Kerr
interaction. (a) The squeezing factor $S$ as a function of evolution time
with $\protect\kappa/\protect\delta=0.95$, $0.99$, $1.0$, $1.05$ as labeled.
The blue solid and orange dashed lines are for $U=10^{-6}\protect\delta$ and
$U=0$ respectively. (b) The squeezing factor as a function of $\protect\kappa%
/\protect\delta$ with evolution time $|\protect\delta|t=30$ (green solid
line) and $|\protect\delta|t=10$ (purple dash-dotted line) with $U=10^{-6}%
\protect\delta$. The inset shows the zoom in near the EP for $|\protect\delta%
|t=30$, with black dashed line showing the $U=0$ result. In all plots, the
results for $U=10^{-6}\protect\delta$ are obtained in the Fock space with a
cutoff photon number of 4000 in each mode, starting from the vacuum state.
The results for $U=0$ are obtained from the analytical solutions of the
field operators.}
\label{fig:Kerr}
\end{figure}

\emph{{\color{blue}Effects of Kerr interaction}.---}Now we discuss the
effects of Kerr interactions~\cite%
{PhysRevA.94.033841S,ncomms11417S,PhysRevA.98.042118S,PhysRevLett.123.173601S,PhysRevLett.125.240405S,PhysRevA.103.033711S,arXiv:2107.04503S}
$\mathcal{H}_\text{Kerr}=\frac{U}{2}\sum_{i}(\hat{a}_{i}^{\dag }\hat{a}%
_{i})^{2}$ on the dynamics near the EP. Kerr interaction strength $U$
corresponds to the nonlinearity at the single-particle level, which is much
smaller than $\kappa $. Our model can be realized by parametric down
conversion and four-wave mixing in quantum optics, as well as cold atomic
BECs in a ring trap. For the parametric down conversion, the typical
second-order nonlinear coefficient is about 10 pm/V, and we have $\kappa =0.5
$ rad/m for typical pumping power $\sim 1$ W. The Kerr nonlinear index
(i.e., third-order nonlinear coefficient) is typically of the order of $%
10^{-18}$ m$^{2}$/W, leading to single-photon Kerr interaction strength $%
\sim 10^{-18}$ rad/m (i.e., $U\sim 10^{-18}\kappa $). For our proposed
four-wave mixing, the ratio $U/\kappa $ is given by the ratio between the
power of a single photon and the pumping laser, which is about $10^{-9}$
(i.e., $U\sim 10^{-9}\kappa $). In the above estimation, we have assumed a
beam waist of about 1 mm and the generated photons have a bandwidth of the
order of MHz. For the cold atomic BEC system, we have $U\simeq \frac{\kappa
}{|\Phi |^{2}}$. The atom number of a BEC can be up to $|\Phi |^{2}\sim
10^{9}$~\cite{PhysRevLett.81.3811S}, which corresponds to $U\sim
10^{-9}\kappa $. For most experiments, the BEC atom number ranges from $%
10^{5}$ to several $10^{6}$~\cite{PhysRep.529.265S}, corresponding to $U\sim
10^{-6}\kappa $.

In Fig.~\ref{fig:Kerr}, we plot the numerical simulations of the squeezing
factor $S$ for $U=10^{-6}\delta $ ($\delta $ and $\kappa $ are of the same
order around EP) with various evolution time $t$ and $\kappa /\delta $. Fig.~%
\ref{fig:Kerr}a shows the squeezing factor as a function of evolution time
with different $\kappa /\delta $. We see that, the oscillation dynamics for $%
\kappa <\delta $ are hardly affected by the Kerr interaction. The linear and
exponential increases of $S$ for $\kappa =\delta $ and $\kappa >\delta $
remain the same when the evolution time is not too long, and the Kerr
interaction would suppress the squeezing as the evolution time and squeezing
factor increase. In our numerical simulation, finite number of Fock states
are considered with a cutoff of 4000 photons in each mode. Numerical errors
become more significant than the Kerr effects when the generated photons $%
\sim S^{2}$ is of the order of 4000, as seen at $|\delta |t>10$ for $\kappa
/\delta =1.05$. Fig.~\ref{fig:Kerr}b shows the squeezing factor as a
function of $\kappa /\delta $ with evolution time $|\delta |t=10$ and $30$.
We see that, the dynamics are hardly affected and squeezing factor still
exhibits sharp changes at the working point of the sensor (the working point
can be up to $\kappa /\delta \simeq 0.995$ for $|\delta |t=30$). Numerical
simulation indicates that the working point of the sensor can be up to $%
\kappa /\delta \simeq 0.995$ even for $U\sim 10^{-6}\kappa $ as in the BEC
systems. The working point can be even closer to the EP point for optical
systems with much weaker Kerr interactions.

\end{widetext}

\end{document}